\documentclass[aps, twocolumn, showpacs, letterpaper]{revtex4}

\pdfoutput=1

\usepackage{amsmath}
\usepackage{amssymb}
\usepackage{graphicx}
\usepackage{xspace}
\usepackage{upgreek}
\usepackage{accents}

\newcommand{\eg}{{e.g.,\/}\xspace}
\newcommand{\ie}{{i.e.,\/}\xspace}

\newcommand{\gap}{\mbox{}}

\newcommand{\eq}[1]{(\ref{#1})}
\newcommand{\Eq}[1]{Eq.~(\ref{#1})}
\newcommand{\Eqs}[1]{Eqs.~(\ref{#1})} 
\newcommand{\Fig}[1]{Fig.~\ref{#1}} 
\newcommand{\Ref}[1]{Ref.~\cite{#1}}
\newcommand{\Refs}[1]{Refs.~\cite{#1}}
\newcommand{\Sec}[1]{Sec.~\ref{#1}}

\newcommand{\App}[1]{Appendix~\ref{#1}}

\newcommand{\mc}[1]{\mathcal{#1}}
\newcommand{\mcc}[1]{\mathfrak{#1}}

\newcommand{\favr}[1]{\langle #1 \rangle}

\newcommand{\pd}{\partial}

\newcommand{\kpt}[1]{{\kern #1 pt}}
\newcommand{\const}{\text{const}}

\begin{document}
\title{Adiabatic nonlinear waves with trapped particles: III. Wave dynamics}
\author{I.~Y. Dodin and N.~J. Fisch}
\affiliation{Department of Astrophysical Sciences, Princeton University, Princeton, New Jersey 08544, USA}
\date{\today}


% 52.35.-g -- Waves, oscillations, and instabilities in plasmas and intense beams
% 52.35.Mw -- Nonlinear phenomena: waves, wave propagation, and other interactions 
%             (including parametric effects, mode coupling, ponderomotive effects, etc.) 
% 52.25.-b -- Plasma properties
% 45.20.Jj -- Lagrangian and Hamiltonian mechanics 

\begin{abstract}
The evolution of adiabatic waves with autoresonant trapped particles is described within the Lagrangian model developed in Paper~I, under the assumption that the action distribution of these particles is conserved, and, in particular, that their number within each wavelength is a fixed independent parameter of the problem. One-dimensional nonlinear Langmuir waves with deeply trapped electrons are addressed as a paradigmatic example. For a stationary wave, tunneling into overcritical plasma is explained from the standpoint of the action conservation theorem. For a nonstationary wave, qualitatively different regimes are realized depending on the initial parameter $S$, which is the ratio of the energy flux carried by trapped particles to that carried by passing particles. At $S < 1/2$, a wave is stable and exhibits group velocity splitting. At $S > 1/2$, the trapped-particle modulational instability (TPMI) develops, in contrast with the existing theories of the TPMI yet in agreement with the general sideband instability theory. Remarkably, these effects are not captured by the nonlinear Schr\"odinger equation, which is traditionally considered as a universal model of wave self-action but misses the trapped-particle oscillation-center inertia.
\end{abstract}

\maketitle
%\bibliographystyle{brief}

%%%%%%%%%%%%%%%%%%%%%%%%%%%%%%%%%%%%%%%%%%%%%%%%%%%%%%%%%%%%
\section{Introduction}
\label{sec:intro}

Within the geometrical-optics (GO) approximation, adiabatic nonlinear waves in collisionless plasma are described conveniently within the average-Lagrangian formalism originally proposed by Whitham \cite{ref:whitham65, book:whitham}. In our \Refs{tex:myacti, tex:myactii}, further called Papers I and II, this formalism was restated to also accommodate effects of autoresonant particles trapped in wave troughs. Specifically, the corresponding Lagrangian density $\mcc{L}$ and the nonlinear dispersion relation (NDR) were derived in Papers I and II (see also \Ref{my:bgk} and references therein), under the assumption that the action distribution of these particles is conserved, and, in particular, that their number within each wavelength is a fixed independent parameter of the problem. Here, the evolution of such waves will be studied.

It is commonly believed that, within the GO approximation, the wave evolution can be inferred from the NDR alone, in generalization of the well-known linear solution \cite{book:karpman}. This implies, in particular, that $\mcc{L}$ is deducible from the NDR. Since the latter is equivalent (Paper~I) to
\begin{gather}\label{eq:la0}
\mcc{L}_a = 0,
\end{gather}
where $a$ is the amplitude \cite{foot:sub}, the assumption hereby is that the integration constant associated with \Eq{eq:la0} is insignificant, so one can take $\mcc{L} = \int^a_0 \mcc{L}_a\,da$. However, this does not apply to waves with autoresonant trapped particles. First of all, at small enough $a$ the adiabatic Lagrangian ceases to exist, because particles start to escape the wave potential. (Only $\delta$-shaped distributions of trapped particles are allowed at $a \to 0$.) Second, as flows from Paper I, $\mcc{L}$ can contain terms which are independent of $a$ [so they cannot be inferred from \Eq{eq:la0}] and yet depend on $\omega$ and $k$, thus affecting the wave dynamics. This is also understood from the fact that trapped particles carry fractions of the wave momentum density $\rho = k \mcc{L}_\omega$ and the energy flux density $\Pi = - \omega \mcc{L}_k$ \cite{book:whitham}, a part of which is determined by the phase velocity rather than $a$. Therefore, examining just the NDR is insufficient to predict the evolution of these waves, contrary to \Refs{ref:ikezi78, ref:dewar72d, ref:rose05, ref:rose08, ref:istomin72b, ref:benisti07, ref:benisti09, ref:benisti10}. Instead, the complete Lagrangian must be used \cite{foot:alt}.

Here, we show how effects captured by $\mcc{L}$ but not by the NDR render the wave self-action due to autoresonant trapped particles unique among other self-action mechanisms. Specifically, we analyze the evolution of one-dimensional (1D) nonlinear Langmuir waves with deeply trapped autoresonant electrons as a paradigmatic example, illustrating the qualitative physics that is also expected for other distributions (except when dynamic trapping and detrapping become essential). For a stationary wave, tunneling into overcritical plasma \cite{ref:krasovsky92, ref:krasovsky92c, ref:krasovsky92b} is explained from the standpoint of the action conservation theorem (ACT). For a nonstationary wave, qualitatively different regimes are realized depending on the initial parameter $S$, which is the ratio of the energy flux carried by trapped particles to that carried by passing particles. At $S < 1/2$, a wave is stable and exhibits group velocity splitting. At $S > 1/2$, the trapped-particle modulational instability (TPMI) develops, in contrast with the existing theories of the TPMI \cite{ref:dewar72d, ref:rose05, ref:rose08} yet in agreement with the general sideband instability (SI) theory \cite{ref:krasovsky94}. Remarkably, these effects are not captured by the nonlinear Schr\"odinger equation (NLSE), which is traditionally considered as a universal model of wave self-action \cite{book:karpman} but misses the inertial of the trapped-particle oscillation centers (OC).

The work is organized as follows. In \Sec{sec:basic}, we formulate our analytical model in general. In \Sec{sec:go}, we present the wave Lagrangian in a simple form and derive the GO equations flowing from it. In \Sec{sec:inhom}, we study waves in plasmas with parameters varying in space and time; in particular, wave tunneling into overcritical plasma is discussed. In \Sec{sec:pulse}, we consider pulse propagation in homogeneous stationary plasma; specifically, the group velocity splitting is derived for $S < 1/2$, and the TPMI rate is calculated for $S > 1/2$. In \Sec{sec:discuss}, we explain our results in the context of a more general stability criterion and compare them with other existing theories. In \Sec{sec:conc}, we summarize our main conclusions. Also, supplementary material is given in appendixes.

%%%%%%%%%%%%%%%%%%%%%%%%%%%%%%%%%%%%%%%%%%%%%%%%%%%%%%%%%%%%
\section{Basic model}
\label{sec:basic}

The assumption that the number of trapped particles is fixed implies that (i) those are trapped in autoresonance \textit{deeply}, such that they do not become untrapped when the wave parameters evolve; (ii) also, it is implied that there are no passing particles close to the resonance, so that no additional trapping can result from the wave evolution. (As models, corresponding distributions already proved useful for understanding paradigmatic effects driven by trapped particles \cite{ref:kruer69, ref:goldman70, ref:goldman71, ref:krasovsky94, ref:krasovsky09}; yet, they can also form naturally as waves evolve \cite{ref:krasovskii95}.) Then, assuming also that the wave envelope is smooth and evolves slowly enough, an adiabatic Lagrangian density $\mcc{L}$ in the GO approximation can be inferred from the general formalism reported in Paper~I. For simplicity, we will focus on 1D nondissipative electrostatic waves in nonmagnetized plasma here. In this case, $\mcc{L}$ reads as (Paper~I)
\begin{gather}\label{eq:mccL0}
\mcc{L} = \frac{(\pd_x \bar{\varphi})^2}{8\pi} + \frac{a^2}{16\pi}  - \sum_s n_s \favr{\mc{H}_s}_{f_s},
\end{gather}
which describes both the quasistatic potential $\bar{\varphi}$ (if any) and also the wave, to be characterized by the amplitude $a$, the frequency $\omega$, and the wave number $k$. The summation in \Eq{eq:mccL0} is taken over different species; $n_s$ are the corresponding densities averaged locally over the wave oscillations; $\mc{H}_s$ are the OC energies, and the angular brackets denote averaging over the particle distributions~$f_s$.

We assume that $\omega$ is large enough, such that no ions are trapped and that the ion quiver motion is insignificant.~Then, 
\begin{gather}
n_i\favr{\mc{H}_i}_{f_i} = n_i\mc{H}_i = \mcc{H}_i  + e n_i \bar{\varphi},
\end{gather}
where $\mcc{H}_i$ is the ion thermal energy density, $e > 0$ is minus the electron charge, and $n_i$ is the ion density, which is equal to the electron unperturbed density $n_0$. (For clarity, we take the ion charge to be $e$.) Suppose also that the number of trapped electrons is small enough so that the wave can be considered monochromatic; for specific conditions see \Refs{ref:krasovsky94, ref:rose01, ref:winjum07}. Then, provided that all passing electrons are nonresonant, thus undergoing \textit{linear} oscillations, one can write (Paper~I)
\begin{gather}
n^{(p)}_e \favr{\mc{H}^{(p)}_e}_{f_p} = \mcc{H}_e - (\epsilon - 1)\,\frac{a^2}{16\pi} - e n^{(p)}_e \bar{\varphi},
\end{gather}
where $\mcc{H}_e$ is the electron thermal energy density, and $\epsilon(\omega, k)$ is the linear dielectric function. Finally, trapped electrons have (Paper~I)
\begin{gather} \label{eq:ht}
 \mc{H}^{(t)}_e = \mc{E}_e^{(t)} - m_e u^2/2,
\end{gather}
where $\mc{E}_e^{(t)}$ is the particle energy in the frame where the wave is stationary, $u = \omega/k$ is the phase velocity, and $m_e$ is the electron mass. Here, we will assume, for simplicity, that these particles initially reside at (and, due autoresonance, stick to) the very bottom of the wave potential troughs, so $\mc{E}_e^{(t)} = - ea/k - e\bar{\varphi}$. Hence, 
\begin{gather}\label{eq:aux111}
n^{(t)}_e \favr{\mc{H}^{(t)}_e}_{f_t} = - en^{(t)}_ea/k - e n^{(t)}_e \bar{\varphi} - m_e n^{(t)}_e u^2/2.
\end{gather}

Combining the above equations yields 
\begin{multline}\label{eq:mccL}
\mcc{L} = \frac{\epsilon a^2}{16\pi} + e \sigma a + \frac{m \sigma \omega^2}{2k}\\
+  \frac{(\pd_x\bar{\varphi})^2}{8\pi} - e\big[n_0 - n^{(p)}\big]\bar{\varphi} + e \sigma k \bar{\varphi}.
\end{multline}
Here we omitted the insignificant thermal energies, dropped the index $e$, and introduced $\sigma = n^{(t)}/k$, which is proportional to the number of trapped electrons within the local wavelength. In this paper, we will assume ${\sigma(x, t) = \const}$ for simplicity; otherwise, see Paper~I.

In what follows, we will also assume that the wave is close to stationary at each $x$. Then, to prevent the quasistatic field build-up, the total electron current must remain approximately zero (assuming that the ion current is zero). Hence, the flow velocity of passing electrons, $V_0$, is estimated as \cite{ref:krasovsky92c, ref:krasovsky92b}
\begin{gather}\label{eq:V0}
 V_0 \sim u n^{(t)}/n^{(p)}. 
\end{gather}
We will assume that such $V_0$ does not affect the plasma dispersion, the condition being that the right-hand side of \Eq{eq:V0} remain small compared to other characteristic velocities in the system. In particular, we will require that it be small compared to the linear group velocity $v_{g0}$. (Remember that $v_{g0} \ll v_T \ll u$, where $v_T$ is the passing-electron thermal velocity; see also \Sec{sec:pulse}.) Thus,
\begin{gather}\label{eq:N}
\mc{N} \equiv \frac{n^{(t)}}{\kappa^2 n^{(p)}} \ll 1
\end{gather}
will be assumed, where
\begin{gather}\label{eq:kappa}
\kappa \equiv k \lambda_D \ll 1,
\end{gather}
and $\lambda_D = v_T/\omega_p$ is the Debye length. Since the bulk plasma is supposed to be cold, one can then take~\cite{my:dense} 
\begin{gather}\label{eq:eps}
\epsilon = 1 - \frac{\omega_p^2}{\omega^2 - 3k^2 v_T^2},
\end{gather}
where $\omega_p = [4\pi n^{(p)}e^2/m]^{1/2}$ is the plasma frequency.

%%%%%%%%%%%%%%%%%%%%%%%%%%%%%%%%%%%%%%%%%%%%%%%%%%%%%%%%%%%%
\section{Wave equations}
\label{sec:go}

Varying $\mcc{L}$ with respect to the wave variables yields wave equations in the geometrical-optics approximation as discussed in Paper~I (see also \App{app:phi}). In particular, $\delta_a \mcc{L} = 0$ yields \Eq{eq:la0}, or
\begin{gather}\label{eq:ndr}
\epsilon(\omega, k) + 8\pi \sigma e/a = 0,
\end{gather}
which defines the wave NDR, $\omega = \omega(k, a)$. Using \Eq{eq:eps}, one obtains then (cf. Paper~I)
\begin{gather}\label{eq:ndr2}
\omega^2 = \frac{\omega_p^2}{1 + 2\omega_t^2/\omega_E^2} + 3 k^2v_T^2,
\end{gather}
where $\omega_t = [4\pi n^{(t)}e^2/m]^{1/2}$, and $\omega_E = (eak/m)^{1/2}$ is the characteristic frequency of the trapped-particle bounce oscillations in the wave troughs. Notice that the sinusoidal-wave approximation requires \cite{ref:krasovsky94}
\begin{gather}\label{eq:vartheta}
\vartheta \equiv \omega_t^2/\omega_E^2 \ll 1.
\end{gather}
Thus, even with the nonlinear corrections, within our model one still has the approximate equality $\omega^2 \approx \omega_p^2$; \ie studied here are weakly nonlinear Langmuir waves. 

Further, notice that $k = \pd_x \xi$ and $\omega = -\pd_t \xi$, where $\xi$ is the wave phase. Then $\delta_\xi \mcc{L} = 0$ gives the ACT,
\begin{gather}\label{eq:act}
\pd_t \mc{I} + \pd_x \mc{J} = 0,
\end{gather}
where $\mc{I} \equiv \mcc{L}_\omega$ is the action density,
\begin{gather}
\mc{I} = \frac{\epsilon_\omega a^2}{16\pi} + m \sigma u, 
\end{gather}
and $\mc{J} \equiv - \mcc{L}_k$ is the action flux density,
\begin{gather}\label{eq:J}
\mc{J} = - \frac{\epsilon_k a^2}{16\pi} + \sigma\left(\frac{m u^2}{2} - e \bar{\varphi}\right).
\end{gather}

As argued in \App{app:phi}, the contribution of $e \bar{\varphi}$ in \Eq{eq:J} can be neglected in the paradigmatic regimes that we consider here. (For effects of prescribed nonzero $\bar{\varphi}$, which also flow from \Eq{eq:act}, see \Ref{ref:matveev08}.) This eliminates $\bar{\varphi}$ from the wave equations altogether, so the wave Lagrangian can henceforth be used in a reduced~form,
\begin{gather}\label{eq:mccLred}
\mcc{L}(a, \omega, k) = \epsilon(\omega, k)\,\frac{a^2}{16\pi} + e \sigma a + \frac{m \sigma \omega^2}{2k},
\end{gather}
yielding, in particular, that
\begin{gather}\label{eq:IJ}
\mc{I} \approx \frac{a^2}{8\pi\omega} + m \sigma u, \quad
\mc{J} \approx \frac{3k v_T^2 a^2}{8\pi \omega^2} + \frac{m \sigma u^2}{2}.
\end{gather}

Remarkably, the contribution of autoresonant trapped particles to $\mcc{L}$ is of \textit{lower} order in $a$ than the leading (first) term. Compared to common nonlinear waves, this is quite unusual. Moreover, the third term in $\mcc{L}$, which is due to the trapped-particle OC inertia, is completely independent of the field amplitude and yet cannot be dropped as it still depends on $\omega$ and $k$. Below, we show how certain paradigmatic effects result from this inertia term, specific to waves with trapped particles. Notice also that, at least qualitatively, those effects are not limited to the model of deeply trapped particles that we adopt. This is because the $a$-independent term in $\mcc{L}$ originates from the second term in \Eq{eq:ht}, which is independent of $\mc{E}^{(t)}_e$ and thus yields a contribution invariant to the trapped-particle distribution. In particular, the application of the Whitham's approach in \Ref{ref:benisti10} is hence invalidated \cite{foot:benisti}, and some other existing theories must be revised, as we will argue in \Sec{sec:discuss}.

%%%%%%%%%%%%%%%%%%%%%%%%%%%%%%%%%%%%%%%%%%%%%%%%%%%%%%%%%%%%
\section{Wave transformations in plasma with varying parameters}
\label{sec:inhom}

Now that $\mcc{L}$ is known explicitly, one can apply the ACT to infer the wave evolution in plasma with parameters varying in space and time. In particular, integrating \Eq{eq:act} over the volume predicts the conservation of the wave total ``number of quanta'', $\int \mc{I}\,dx = \const$. Combined with the NDR, for a homogeneous wave this yields, for instance, $a = a(\omega(t))$ in the same manner as for linear waves discussed in \Refs{my:dense, my:langact, my:mquanta}. Below, we consider another example, namely, stationary wave propagation in inhomogeneous plasma. Unlike in \Refs{ref:krasovsky92, ref:krasovsky92b, ref:krasovsky92c}, where an \textit{ad hoc} solution of the Vlasov equation was employed to address a similar problem, we will show that our ACT yields the same results straightforwardly.

\begin{figure}
\centering
\includegraphics[width=.45\textwidth]{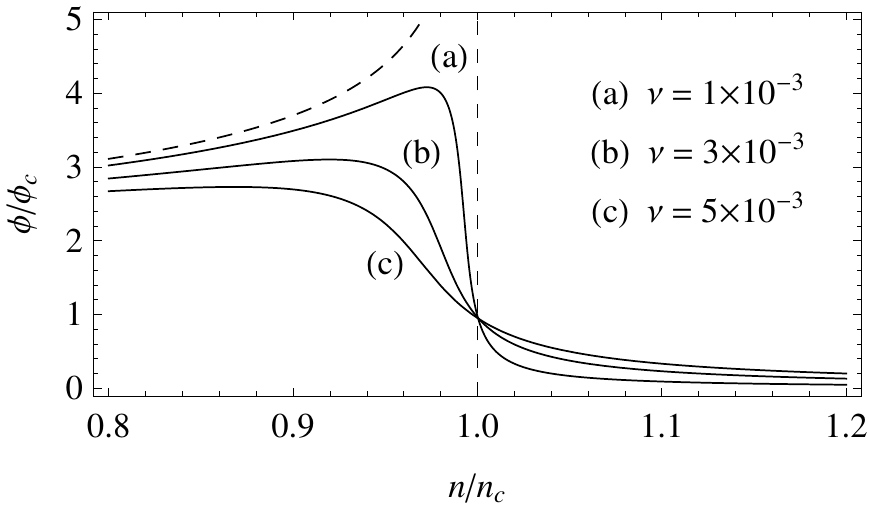}
\caption{1D stationary wave with trapped particles in inhomogeneous plasma. Shown is the normalized amplitude of the wave potential, $\phi/\phi_c$, vs. the plasma normalized density $n/n_c \equiv \Omega_p^2$ for different $\nu$. Here $\phi_c = 2\Theta$, with $\Theta = 0.1$ taken as an example; $n_c$ is the critical density. Dashed is the linear solution ($\nu = 0$) and the location of the linear cutoff ($n = n_c$).}
\label{fig:tunnel}
\end{figure} 

For a stationary wave \Eq{eq:act} gives $\mc{J} = \const$. Since one also has $\pd_x \omega = 0$ in that case [\Eq{eq:cons}], this is understood as conservation of the wave energy flux ${\Pi = \omega \mc{J}}$ (cf. \App{app:vg}),~or
\begin{gather}\label{eq:efl}
\Pi^{(p)} + \Pi^{(t)} = \const,
\end{gather}
where we introduced the densities of energy fluxes carried by passing particles and trapped particles, respectively:
\begin{gather}\label{eq:eflaux}
\Pi^{(p)} = \frac{3k v_T^2 a^2}{8\pi \omega}, \quad \Pi^{(t)} = \frac{1}{2}\,n^{(t)}mu^3
\end{gather}
(cf. also \Refs{ref:krasovsky92c, book:stix}). If the ratio
\begin{gather}\label{eq:S}
S \equiv \Pi^{(t)}/\Pi^{(p)}
\end{gather}
is small, one can anticipate that the wave is in the linear regime and is not affected significantly by trapped particles. However, $S$ can become large as the plasma density varies along the ray trajectory. Then the wave dynamics becomes essentially nonlinear, which is seen as follows.

Let us introduce $\Omega_p(x) = \omega_p(x)/\omega$ and $\varkappa = k \lambda_T \approx \kappa$, where $\lambda_T = v_T/\omega \approx \lambda_D$; also, $\nu = \sigma/(n_c \lambda_T) \approx \mc{N}$, with $n_c$ being the critical density, and $\phi = e a/(m\omega v_T)$. Then \Eq{eq:efl} rewrites as
\begin{gather}\label{eq:prec}
3\varkappa \phi^2 + \nu/\varkappa^2 = 3\Theta,
\end{gather}
where $\Theta$ is some constant determined by boundary conditions. Also, we can substitute $\varkappa$ from \Eq{eq:ndr2}, namely,
\begin{gather}\label{eq:gold}
3\varkappa^2 \approx (1 - \Omega_p^2) + 2\nu/\phi.
\end{gather}
Then the equation for $\phi$ reads as
\begin{gather}\label{eq:efl2}
\frac{\phi^2}{\sqrt{3}}\,\sqrt{1 - \Omega_p^2 + 2\nu/\phi} + \frac{\nu}{1 - \Omega_p^2 + 2\nu/\phi} = \Theta.
\end{gather}
Hence, the effect of trapped particles is determined by two parameters: $S$, defined [in agreement with \Eq{eq:S}] as the ratio of the terms on the left-hand side here, and also $\varsigma$, which is a measure of how $\varkappa$ is affected by the nonlinearity [cf. \Eq{eq:gold}]; namely,
\begin{gather}
S \sim \frac{\nu/\varkappa^2}{\varkappa \phi^2}, \quad
\varsigma \equiv \frac{\nu/\phi}{\varkappa^2}.
\end{gather}

Suppose that a wave is launched from a subcritical region ($\Omega_p < 1$) and initially is close to linear ($S \ll 1$, $\varsigma \ll 1$), the boundary condition thus being $\Theta \approx \varkappa_0 \phi_0^2$. Then, at first, $\phi \approx \phi_0 (\varkappa_0/\varkappa)^{1/2}$, so $\varsigma$ remains constant. Yet $S$ grows as $\varkappa^{-2}$, and eventually one gets $S \sim 1$, at some $x = x_*$ \cite{foot:xstar}. Beyond that point, the second term in \Eq{eq:prec} dominates, resulting in constant $\varkappa^2 \approx \nu/(3\Theta)$. Hence \Eq{eq:gold} yields $\Omega_p^2 - 1 + \nu/\Theta= 2\nu/\phi$,~or
\begin{gather}\label{eq:aux112}
\phi = \frac{2\nu}{\Omega_p^2 - 1 + \nu/\Theta}.
\end{gather}
As $\Omega_p$ grows further, $\phi$ starts to \textit{decrease} then (\Fig{fig:tunnel}). Finally, the wave reaches $x = x_c$, where $\Omega_p^2 = 1$ [and thus ${\varsigma(x_c) \sim 1}$]. At that location, $\phi$ equals about $2\Theta$ and then continues analytically into the overcritical region. 

Hence, in contrast with a linear wave (for which a GO solution with nonzero $\Pi$ would be impossible there), a nonlinear wave loaded with trapped particles can, in principle, penetrate overcritical plasma, in agreement with \Refs{ref:krasovsky92, ref:krasovsky92b, ref:krasovsky92c}. In particular, since the inhomogeneity scale does not enter the above equations explicitly, it can always be chosen large enough, so as to ensure the validity of the GO profile we derived. However, notice that such a wave will propagate adiabatically for limited time only. This is because $S(x \gtrsim x_c) \gtrsim (\nu \Theta)^{-1/2} \gg 1$, in which case the wave is unstable, as we will now discuss.

\begin{figure}[t]
\centering
\includegraphics[width=.5\textwidth]{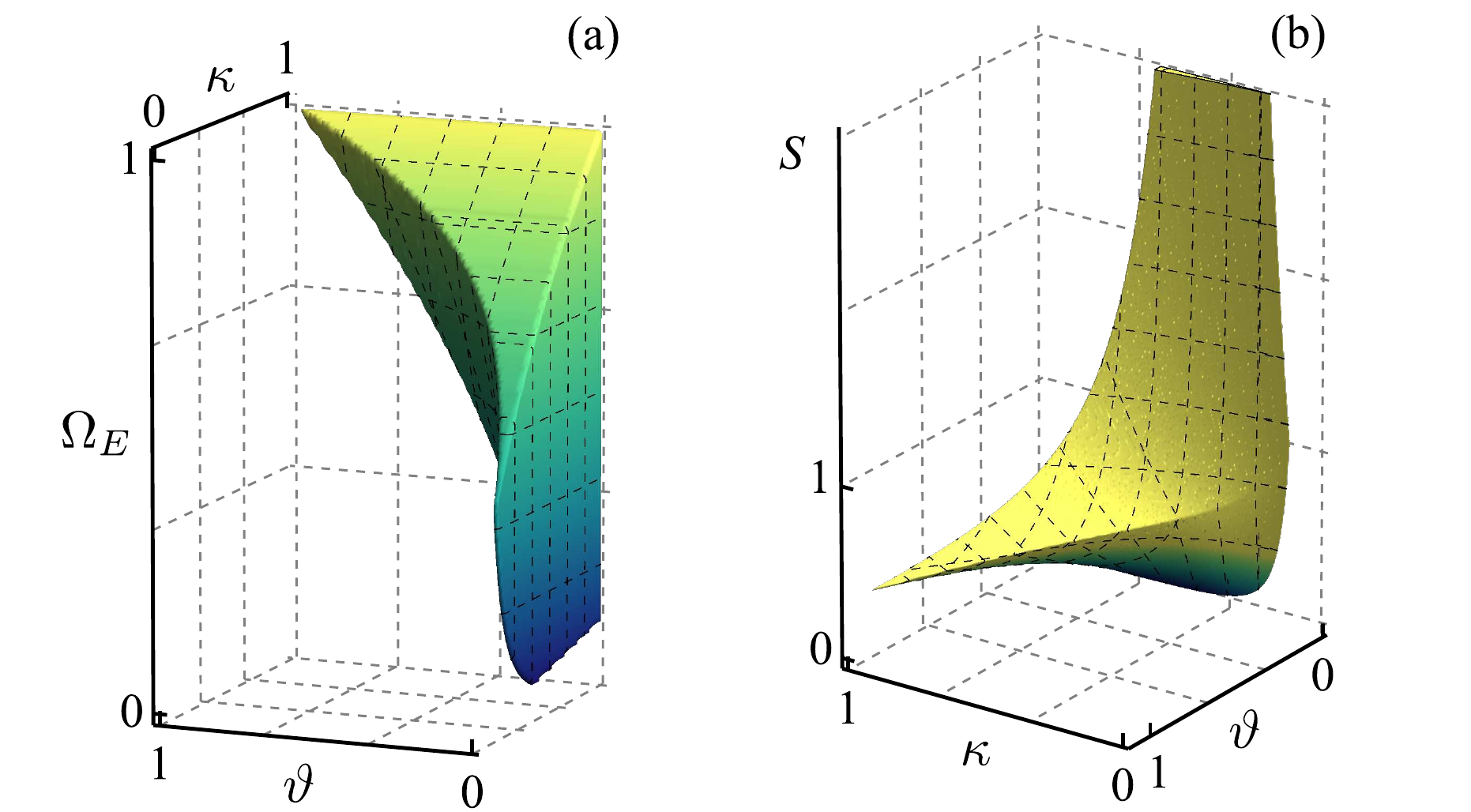}
\caption{(Color online) Schematic of the parameter domain assumed for \Sec{sec:pulse}: (a) in space ($\kappa, \vartheta, \Omega_E$), (b) in space ($\kappa, \vartheta, S$). Combined here are the following assumptions: the plasma is cold [\Eq{eq:kappa}], the wave is sinusoidal [\Eq{eq:vartheta}] and weak enough [\Eq{eq:PhiTineq}], the quasistatic field due to trapped particles is negligible [\Eq{eq:Sk4}]; also, the bulk motion and the nonlinear effects are weak, \ie $V_0 \ll \Delta v_g \ll v_{g0}$ [\Eq{eq:dvineq}; see also \Eqs{eq:V0} and \eq{eq:dv}]. The inequalities~\eq{eq:N} and \eq{eq:PhiJ} reduce to $V_0 \ll v_{g0}$ and thus are satisfied automatically.}
\label{fig:domain}
\end{figure}

%%%%%%%%%%%%%%%%%%%%%%%%%%%%%%%%%%%%%%%%%%%%%%%%%%%%%%%%%%%%
\section{Pulse propagation}
\label{sec:pulse}

Let us consider the envelope dynamics, assuming, for simplicity, that the plasma is homogeneous and stationary. Since the ACT [\Eq{eq:act}], serving as the envelope equation, contains only first-order derivatives of $a$, the diffraction is hereby neglected. Within this approximation, a linear pulse would conserve its shape, traveling at the linear group velocity $v_{g0}$. A nonlinear wave, in contrast, will undergo distortion, particularly because it can have \textit{two} group velocities~$v_g$.

%-----------------------------------------------------------
\subsection{Group velocity splitting}
\label{sec:gvsp}

\begin{figure}[b]
\centering
\includegraphics[width=.48\textwidth]{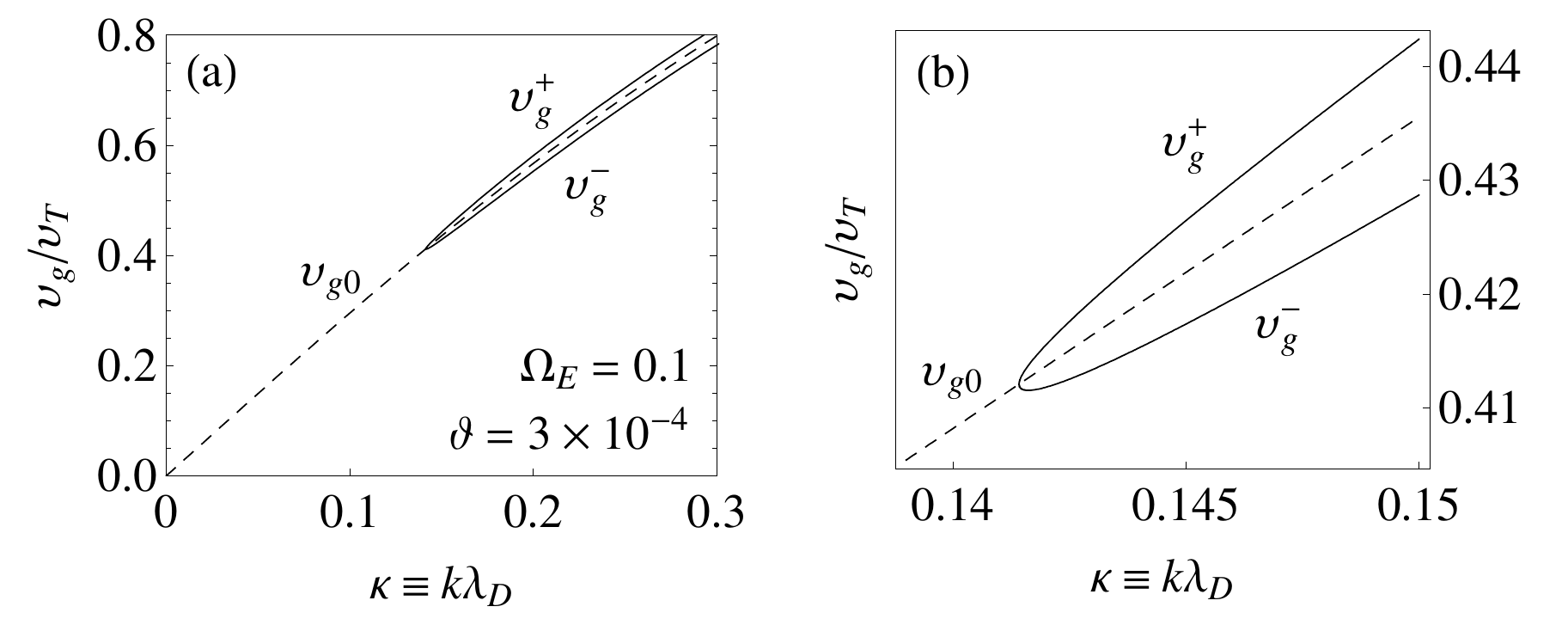}
\caption{(a) Nonlinear group velocities, $v_g^+$ and $v_g^-$ [\Eq{eq:dv2}] vs. $\kappa$, for sample $\Omega_E$ and $\vartheta$; the dashed line shows the linear group velocity $v_{g0}$. At $S > 1/2$, corresponding to $\kappa < \kappa_S \equiv \Omega_E^{-1}\sqrt{2\vartheta/3}$, no real solutions exist for $v_g$, rendering the wave unstable. (b) Close-up at $\kappa \approx \kappa_S$.}
\label{fig:vg}
\end{figure}

Specifically, following \Refs{ref:whitham65, book:whitham}, we define $v_g$ here as the velocities of information (rather than that of the energy, like in \Refs{ref:lighthill65, ref:decker94, ref:decker95, ref:schroeder11}), which propagates along characteristics of the wave equation (\App{app:vg}). Hence, $v_g$ are found explicitly from \Eqs{eq:quadr}-\eq{eq:q}, via substituting \Eq{eq:mccLred} for $\mcc{L}$. Using \Eqs{eq:kappa} and \eq{eq:vartheta}, after a tedious yet straightforward calculation, one gets~then
\begin{gather}\label{eq:dv}
v_g = v_{g0}\,\frac{1 + S\vartheta \pm g}{1 + 3 S\vartheta \kappa^2},
\end{gather}
where $v_{g0} \approx 3 \kappa v_T$, and
\begin{gather}
g = \Omega_E \sqrt{S \left(1/2 - S\right)}, 
\end{gather}
with $\Omega_E \equiv \omega_E/\omega_p$; in particular, $S = \vartheta/(3\Omega_E^2 \kappa^2)$. We assume that the nonlinear corrections to $v_{g0}$ are small,~\ie
\begin{gather}\label{eq:dvineq}
S\vartheta \ll 1, \quad g \ll 1,
\end{gather}
where the former inequality ensures that the denominator in \Eq{eq:dv} is also close to unity. (For a summary of our assumptions, see \Fig{fig:domain}.) Hence,
\begin{gather}\label{eq:dv2}
v_g \approx v_{g0} \Big[1 \pm \Omega_E \sqrt{S \left(1/2 - S\right)}\,\Big],
\end{gather}
with a characteristic shape of $v_g(\kappa)$ shown in \Fig{fig:vg}. 

\begin{figure}
\centering
\includegraphics[width=.48\textwidth]{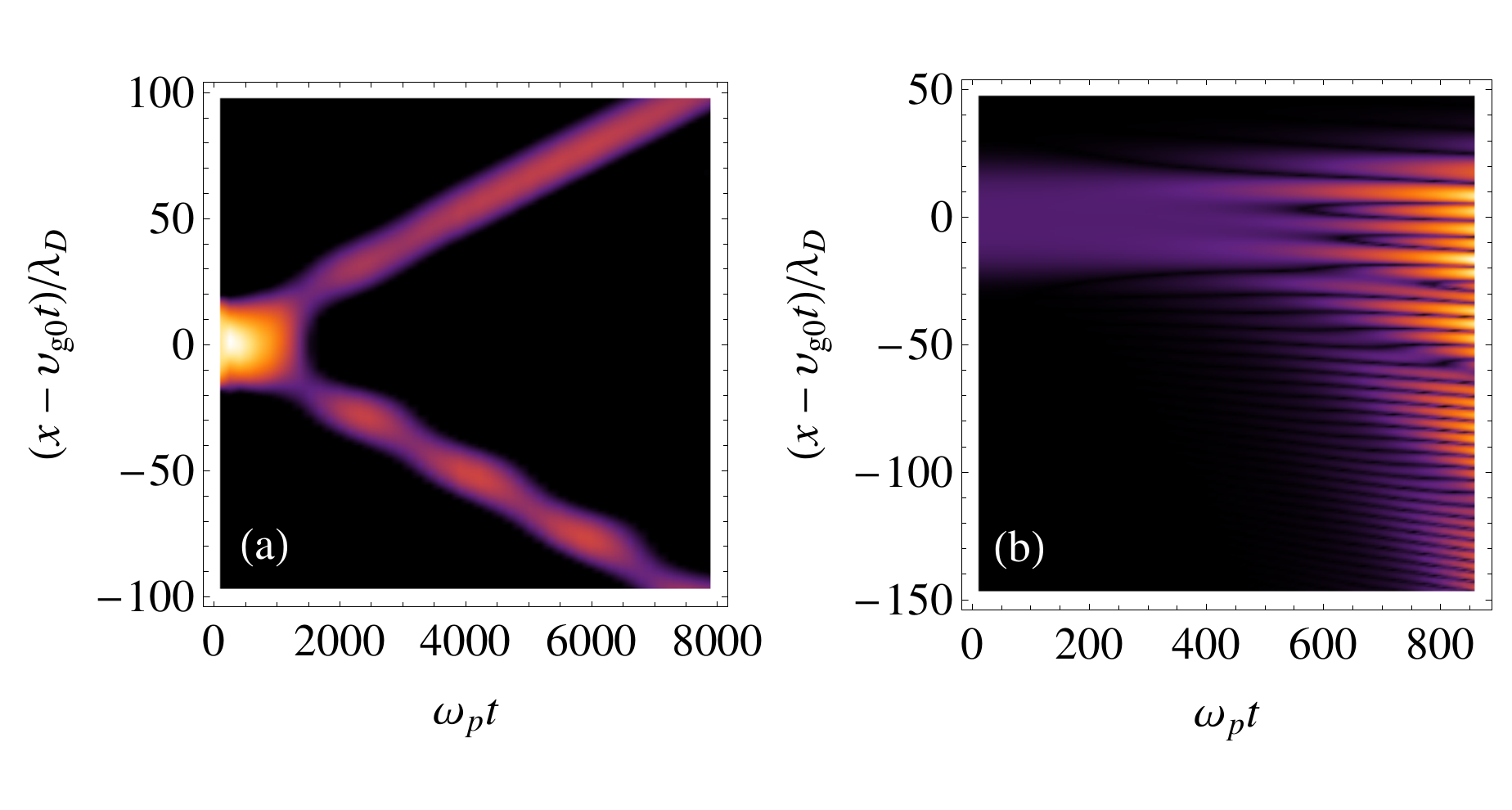}
\caption{(Color online) Evolution of the perturbation $\Delta (a_0^2) = - 0.03 a_0^2 \exp(-x^4/\ell^4)$ to a homogeneous wave with the initial amplitude $a_0$. The solution is obtained by numerical integration of \Eqs{eq:la}-\eq{eq:actwith}, with $\mcc{L}$ taken from \Eq{eq:mccLred}, for the same parameters as in \Fig{fig:vg} and $\ell = 20 \lambda_D$. Shown is $\Delta (a^2)$ (arbitrary color scaling), vs. $t$ and $x$ in the frame moving with the linear group velocity $v_{g0}$; the units are $\omega_p^{-1}$ and $\lambda_D$, correspondingly. (a)~$\kappa = 0.2$, so $S = 1/4$; the wave is stable, resulting in signal splitting. (b)~$\kappa = 0.1$, so $S = 1$; the wave is TPMI-unstable.}
\label{fig:propag}
\end{figure}

Although small, the second term in \Eq{eq:dv2} is retained because it is responsible for essentially nonlinear effects. In particular, consider the case when $S < 1/2$. Then $g$ is real, causing the group velocity splitting by $\Delta v_g = 2v_{g0}g$. This means, for example, that a general modulation imposed on the wave profile eventually splits into two signals propagating with different velocities [\Fig{fig:propag}(a)]. Each signal may then evolve further, if having a finite spread of $a$ and thus of $v_g$ too; however, such a signal will be comprised of characteristics that all correspond to the same sign in \Eq{eq:dv2}, so further splitting \textit{per~se} will not occur. As we remind in \App{app:vg}, the pulse splitting is an inherent feature of all nonlinear waves, as well known in classical hydrodynamics \cite{book:whitham, book:landau6} and also observed in plasma physics experiments \cite{ref:kunhardt76, ref:ikezi78}.

%-----------------------------------------------------------
\subsection{TPMI}
\label{sec:tpmi}

In contrast, at $S > 1/2$, \Eq{eq:dv2} yields no real solutions for $v_g$. Then, the wave is TPMI-unstable, as illustrated in \Fig{fig:propag}(b). One can also assess the wave stability using another approach, which yields the TPMI rate explicitly. Instead of searching for characteristics, take
\begin{gather}
a = a_0 + \Delta a, \quad
\omega = \omega_0 + \Delta \omega, \quad
k = k_0 + \Delta k
\end{gather}
and assume that the perturbations, denoted by $\Delta$, are small compared to the corresponding homogeneous parameters, denoted by $0$. Then, one can linearize the GO equations [\Eqs{eq:la}-\eq{eq:actwith}] and find the dispersion relation $\Delta \omega(\Delta k)$, which is linear in the sense that it is independent of $\Delta a$ (albeit not of $a_0$). Specifically, under the same assumptions as for \Eq{eq:dv2}, one can show~that
\begin{gather}\label{eq:dw}
\Delta \omega = \Delta k\,v_{g0}(1 \pm g).
\end{gather}

At $S < 1/2$, one has real $\Delta \omega$, and the signal velocity $v_g$ that we introduced earlier is recovered as the effective \textit{linear} group velocity
\begin{gather}
\frac{d(\Delta \omega)}{d(\Delta k)} = v_{g0}(1 \pm g).
\end{gather}
At $S > 1/2$, one has two $\omega$ complex conjugate to each other. Then the wave is unstable, as predicted earlier; yet now we get a formula for the TPMI rate too,
\begin{gather}\label{eq:gamma1}
\gamma \approx \Delta k\, v_{g0}\Omega_E \sqrt{S \left(S - 1/2\right)}.
\end{gather}

%%%%%%%%%%%%%%%%%%%%%%%%%%%%%%%%%%%%%%%%%%%%%%%%%%%%%%%%%%%%
\section{Discussion}
\label{sec:discuss}

%-----------------------------------------------------------
\subsection{Stability criterion}

The TPMI threshold also can be inferred from the generalized Lighthill's criterion (GLC), which states that a wave is stable when $\omega_{\mc{I}}(k,\mc{I})\mc{J}_k(k,\mc{I}) > 0$ (\App{app:vg}). To see this, let us first rewrite $\mc{I}$ as
\begin{gather}\label{eq:I2}
\mc{I} \approx \frac{a^2}{8\pi\omega_p} + \frac{m \sigma \omega_p}{k},
\end{gather}
from where one gets $a = a(k, \mc{I})$, with $a_\mc{I} > 0$. Since $\omega_a(k, a) > 0$, the latter yields $\omega_{\mc{I}}(k, \mc{I}) > 0$, so the GLC takes the form $\mc{J}_k(k,\mc{I}) > 0$. Further, using that
\begin{gather}\label{eq:J2}
\mc{J} \approx \frac{3k v_T^2 a^2}{8\pi \omega_p^2} + \frac{m \sigma \omega_p^2}{2k^2},
\end{gather}
we obtain
\begin{gather}
\mc{J}_k \approx \frac{3 v_T^2 a^2}{4\pi \omega_p^2}\left(\frac{1}{2} - S + \frac{k a_k}{a}\right).
\end{gather}
Since $k a_k/a \approx 3 S \kappa^2 \ll S$ [from \Eq{eq:I2}], the GLC hereby rewrites as $S < 1/2$, in agreement with the criterion that we found earlier from \Eq{eq:dv2}. 

Notice that waves with other distributions of trapped electrons also can be studied similarly. (Of course, $\sigma$ still must be conserved, which is not the case, \eg for distributions nonzero near the trapping boundary; cf. Paper~II.) For example, if $\omega_a(k, a) < 0$, the GLC would read as $S > 1/2$, assuming that \Eqs{eq:I2} and \eq{eq:J2} still hold approximately. However, remember that the GLC, and the underlying Whitham's formalism overall, neglects diffraction and assesses the stability of only adiabatic perturbations with small enough $\Delta k$. Even when those are stable, at larger $\Delta k$ a more general SI \cite{ref:kruer69, ref:goldman70, ref:goldman71, ref:brunner04} may develop \cite{ref:krasovsky94}, for which the TPMI, when present, can be considered as the adiabatic limit. In particular, notice that it is the general SI rather than the TPMI that is responsible for the effects reported in \Ref{ref:brunner04}.

%-----------------------------------------------------------
\subsection{Comparison with the existing theories}

The SI is treated systematically in \Ref{ref:krasovsky94} from where a rate can be inferred that matches our \Eq{eq:gamma1} under the appropriate conditions (\App{app:kras}). Yet, our results are in drastic variation with the traditional \cite{foot:know} models of the TPMI \cite{ref:dewar72d, ref:rose05, ref:rose08}, for those predict that a wave is always stable when resonant electrons remain deeply trapped~\cite{ref:dewar72d}. The discrepancy is due to the fact that \Refs{ref:dewar72d, ref:rose05, ref:rose08} rely on the NLSE, which neglects the contribution of the $a$-independent term in $\mcc{L}$ and thus is generally inapplicable to waves with trapped particles \cite{foot:nlse}. A more detailed comparison would require accounting for diffraction (which is out of the scope of our model), but it is already seen that the NLSE-based models do not apply at large $S$. (Interestingly, disagreement between those models and numerical simulations has already been reported before; see, \eg \Refs{ref:rose08, ref:masson10}.)

In connection to this, notice that at $\kappa \gtrsim 0.2$, which is usually associated with large enough $n^{(t)}$ \cite{ref:rose01}, $S > 1/2$ is achieved easily, not to mention that SI may develop too. Thus, the effect of the trapped-particle OC inertia on $\gamma$ that we discuss here is important for assessing the wave stability under practical conditions. Also notice that the actual NDR may be nonlocal (Paper~II), contrary to the tacit assumption used commonly. Hence, further investigation may be needed to adequately model the TPMI and related effects.

%%%%%%%%%%%%%%%%%%%%%%%%%%%%%%%%%%%%%%%%%%%%%%%%%%%%%%%%%%%%
\section{Conclusions}
\label{sec:conc}

In this paper, the evolution of adiabatic waves with autoresonant trapped particles is analyzed under the assumption that the action distribution of these particles is conserved, and, in particular, that their number within each wavelength is a fixed independent parameter of the problem. Particularly, 1D nonlinear Langmuir waves with deeply trapped autoresonant electrons are addressed as a paradigmatic example, illustrating the qualitative physics that is also expected for other distributions (except when dynamic trapping and detrapping become essential). For a stationary wave, tunneling into overcritical plasma is explained from the standpoint of the action conservation theorem. For a nonstationary wave, qualitatively different regimes are realized depending on the initial parameter $S$, which is the ratio of the energy flux carried by trapped particles to that carried by passing particles. At $S < 1/2$, a wave is stable and exhibits group velocity splitting. At $S > 1/2$, the trapped-particle modulational instability develops, in contrast with the existing theories of the TPMI yet in agreement with the general SI theory.  Remarkably, these effects are not captured by the nonlinear Schr\"odinger equation, which is traditionally considered as a universal model of wave self-action but misses the trapped-particle OC inertia.

%%%%%%%%%%%%%%%%%%%%%%%%%%%%%%%%%%%%%%%%%%%%%%%%%%%%%%%%%%%%
\section{Acknowledgments}

The work was supported through the NNSA SSAA Program through DOE Research Grant No. DE274-FG52-08NA28553.

\appendix

%%%%%%%%%%%%%%%%%%%%%%%%%%%%%%%%%%%%%%%%%%%%%%%%%%%%%%%%%%%%
\section{Electrostatic potential}
\label{app:phi}

Let us discuss the specific conditions under which the electrostatic potential $\bar{\varphi}$ can be neglected in \Eq{eq:J}. First of all, the equation for $\bar{\varphi}$ is obtained from $\delta_{\bar{\varphi}}\mcc{L} = 0$ [\Eq{eq:mccL}] and represents the usual Poisson's equation,
\begin{gather}\label{eq:poisson}
\pd^2_{xx} \bar{\varphi} = - 4\pi e\big[n_0 - n^{(p)} - n^{(t)}\big],
\end{gather}
where $n_0$ is determined by initial conditions, and $n^{(t)}$ is ``frozen'' into the wave, due to $\sigma = \const$. To find $n^{(p)}$, consider the momentum equation for passing electrons,
\begin{gather}\label{eq:momentum}
m\, (\pd_t V_0 + V_0\,\pd_x V_0) = \pd_x(e\bar{\varphi} - \Phi) - T\,\pd_x \ln n^{(p)},
\end{gather}
where we assumed, for simplicity, that the plasma is isothermal and included the ponderomotive potential, $\Phi \approx e^2 a^2/(4 m\omega_p^2)$ (in contrast with \Ref{ref:krasovsky92c}, where $\Phi$ was left out). 

Since $\Phi/mV_0^2 \sim \vartheta^{-2} \gg 1$, we can neglect the convective term on the left-hand side immediately. The other term can be estimated from
\begin{gather}
m\, \pd_t V_0 \sim m v_{g0}\, \pd_x \tilde{V}_0,
\end{gather}
with tilde henceforth denoting perturbations to stationary values. From the continuity equation,
\begin{gather}
\pd_t n^{(p)} + \pd_x \big[ n^{(p)} V_0 \big] = 0,
\end{gather}
one gets that $\Delta V_0 \sim v_{g0} \tilde{n}^{(p)}/n^{(p)}$; hence,
\begin{gather}
m\, \pd_t V_0 \sim m v_{g0}^2\, \pd_x  \tilde{n}^{(p)}/n^{(p)} \ll T\, \pd_x  \tilde{n}^{(p)}/n^{(p)}.
\end{gather}
This allows one to neglect the inertia term compared to the pressure term in \Eq{eq:momentum}, thus yielding
\begin{gather}\label{eq:static}
\pd_x \big[ - e\bar{\varphi} + \Phi + T\,\ln n^{(p)} \big] = 0.
\end{gather}

For weak enough waves, with $\Phi \ll T$, or 
\begin{gather}\label{eq:PhiTineq}
\Omega_E \ll \kappa,
\end{gather}
one can thereby estimate the maximum $\bar{\varphi}$ as
\begin{gather}\label{eq:phiT}
e\bar{\varphi} \lesssim T \ll mu^2.
\end{gather}
For the purpose of \Sec{sec:inhom}, \Eq{eq:phiT} is already enough to neglect $e\bar{\varphi}$ in \Eq{eq:J}. For the other regimes that we discuss, one may assume $e\bar{\varphi} \ll T$, yet what matters in \Eq{eq:J} now are only \textit{variations} of $mu^2$, which may not be smaller than $e\bar{\varphi}$; thus, \Eq{eq:phiT} becomes insufficient. In this case, let us revert to \Eq{eq:poisson} and substitute 
\begin{gather}
n^{(p)} = n^{(p)}_0 e^{(e\bar{\varphi} - \Phi)/T} \approx n^{(p)}_0 \big[1 + (e\bar{\varphi} - \Phi)/T\big]
\end{gather}
from \Eq{eq:static}; hence,
\begin{gather}\label{eq:poisson2}
\lambda_D^2\,\pd^2_{xx} (e\bar{\varphi}) = e\bar{\varphi} - \Phi + T\,\tilde{n}^{(t)}/n^{(p)}_0,
\end{gather}
where $n^{(p)}_0$ is the unperturbed density of passing electrons. The latter term in \Eq{eq:poisson2} is negligible, due~to
\begin{gather}\label{eq:Sk4}
\frac{T}{\Phi}\,\frac{\tilde{n}^{(t)}}{n^{(p)}_0} \lesssim S \kappa^4 \ll 1,
\end{gather}
provided that $S$ is moderate. (Remember that we are mostly interested in regimes with $S \sim 1$.) At $\lambda_D^2\,\pd^2_{xx} \ll 1$, one gets then $e\bar{\varphi} \approx \Phi$. On the other hand,
\begin{gather}\label{eq:PhiJ}
\frac{\sigma \Phi}{\epsilon_k a^2/16\pi} \sim \mc{N} \ll 1
\end{gather}
[\Eq{eq:N}]. Thus, again, $e\bar{\varphi}$ is negligible in \Eq{eq:J}.

%%%%%%%%%%%%%%%%%%%%%%%%%%%%%%%%%%%%%%%%%%%%%%%%%%%%%%%%%%%%
\section{Nonlinear group velocities and wave stability}
\label{app:vg}

In this appendix, we concisely restate, to avoid ambiguity, the concept of the group velocity for linear and nonlinear waves, as introduced originally by Whitham \cite{ref:whitham65, book:whitham} (see also \Refs{ref:whitham65a, ref:lighthill65, ref:lighthill65b, ref:lighthill67, ref:hayes73}), and discuss how it is connected with the wave stability.

%-----------------------------------------------------------
\subsection{Conservation laws and flow velocities}
\label{app:sflow}

For simplicity, we will assume a 1D system, the generalization to the case of multiple dimensions being straightforward. In the GO approximation, the wave is completely described by its amplitude $a$, frequency $\omega$, and the wave number $k$, so the Lagrangian density can be taken in the form $\mcc{L}(a, \omega, k; t, x)$. Then, like in \Sec{sec:go}, the wave equations read~as
\begin{gather}
\mcc{L}_a = 0, \label{eq:la}\\
\pd_t k + \pd_x \omega = 0, \label{eq:cons}\\
\pd_t \mcc{L}_\omega - \pd_x \mcc{L}_k = 0. \label{eq:actwith}
\end{gather}
Equation \eq{eq:la} defines the NDR, $\omega = \omega(k, a)$. Equation \eq{eq:cons} is the consistency condition, satisfied due to 
\begin{gather}\label{eq:xikom}
\omega = - \pd_t \xi, \quad k = \pd_x \xi,
\end{gather}
where $\xi$ is the field phase. [Notice, in particular, that \Eq{eq:cons} can be understood as the continuity equation for wave crests, with $k$ being the crest density, and $\omega = ku$ being the crest flux density.] Finally, \Eq{eq:actwith} is the ACT, also known in the form \eq{eq:act}, or
\begin{gather}\label{eq:actv}
\pd_t \mc{I} + \pd_x (v_{\mc{I}} \mc{I}) = 0,
\end{gather}
with $v_{\mc{I}} = - \mcc{L}_k/\mcc{L}_\omega$ serving as the action flow velocity.

Suppose now that the medium is stationary, \ie $\mcc{L}$ does not depend on $t$ explicitly. Then, $\pd_t \mcc{L} = \mcc{L}_\omega\,\pd_t \omega + \mcc{L}_k\,\pd_t k$, where we used \Eq{eq:la}. Thus,
\begin{align}
\pd_t (\omega \mcc{L}_\omega - \mcc{L}) 
 & = \omega\, \pd_t \mcc{L}_\omega + \mcc{L}_\omega\, \pd_t \omega - \mcc{L}_\omega\,\pd_t \omega - \mcc{L}_k\,\pd_t k \notag \\
 & = \omega\, \pd_t \mcc{L}_\omega - \mcc{L}_k\,\pd_t k \notag \\
 & = \omega\, \pd_x \mcc{L}_k - \mcc{L}_k\,\pd_t k \notag \\
 & = \omega\, \pd_x \mcc{L}_k + \mcc{L}_k\,\pd_x \omega \notag  \\
 & = \pd_x (\omega \mcc{L}_k), \label{eq:wec}
\end{align}
where \Eqs{eq:cons} and \eq{eq:actwith} were employed. Similarly, if $\mcc{L}$ does not depend on $x$ explicitly, one gets 
\begin{gather}\label{eq:pec}
\pd_x (k \mcc{L}_k - \mcc{L}) = \pd_t (k \mcc{L}_\omega).
\end{gather}
Equations \eq{eq:wec} and \eq{eq:pec} represent the conservation laws for the wave energy and momentum and can be written~as
\begin{gather}
\pd_t \varepsilon + \pd_x (v_\varepsilon \varepsilon) = 0, \quad
\pd_t \rho + \pd_x (v_\rho \rho) = 0,
\end{gather}
where we introduced
\begin{gather}\label{eq:swdef1}
\varepsilon = \omega \mcc{L}_\omega - \mcc{L}, \quad \rho = k \mcc{L}_\omega
\end{gather}
for the energy density and the momentum density and 
\begin{gather}
v_\varepsilon = - \frac{\omega \mcc{L}_k}{\omega \mcc{L}_\omega - \mcc{L}}, \quad 
v_\rho = -\frac{k \mcc{L}_k - \mcc{L}}{k \mcc{L}_\omega}
\end{gather}
for the corresponding flow velocities. In particular, notice that the energy flux density $\Pi \equiv v_\varepsilon \varepsilon$ and the momentum flux density $\mc{P} \equiv v_\rho \rho$ are then given by
\begin{gather}\label{eq:swdef2}
\Pi = - \omega \mcc{L}_k , \quad \mc{P} = \mcc{L} - k \mcc{L}_k.
\end{gather}

In general, $v_{\mc{I}}$, $v_\varepsilon$, and $v_\rho$ are all different from each other. The exception is the linear regime, which is defined as the regime when $\omega(k)$, inferred from \Eq{eq:la}, is independent of $a$. The latter is possible if $\mcc{L}_a$ has the form $\mcc{L}_a = \mcc{D}(\omega, k)A_a$, where $A$ is some function such that $A_a$ is nonzero; hence $\mcc{L} = \mcc{D}(\omega, k)A$. Since \Eq{eq:la} thereby reads as $\mcc{D}(\omega, k) = 0$, one has $\mcc{L} = 0$,~so
\begin{gather}
v_\varepsilon = v_\rho = v_{\mc{I}} = - \mcc{L}_k/\mcc{L}_\omega.
\end{gather}
From differentiating $\mcc{L}(a, \omega(k), k) = 0$ with respect to $k$, one gets $- \mcc{L}_k/\mcc{L}_\omega = \omega_k \equiv v_{g0}$, with the latter known as the linear group velocity. Therefore, in the linear regime,
\begin{gather}\label{eq:vels}
v_\varepsilon = v_\rho = v_{\mc{I}} = v_{g0}.
\end{gather}
Notice also that a pulse is usually linear at its front and tail (except in the presence of trapped particles), since the field is weak there. Hence, it is only within the pulse that \Eq{eq:vels} can be violated, due to nonlinear effects.

%-----------------------------------------------------------
\subsection{Nonlinear group velocity}
\label{app:nvgwith}

By analogy with the linear case [\Eq{eq:vels}], the nonlinear group velocity $v_g$ is often defined as $v_\varepsilon$ too \cite{ref:lighthill65, ref:decker94, ref:decker95, ref:schroeder11}, or as $\omega_k$ with the derivative taken at fixed $\mcc{L}/\omega$, since
\begin{gather}\label{eq:vgen}
(\omega_k)_{\mcc{L}/\omega} = - \frac{\pd_{k} (\mcc{L}/\omega)}{\pd_\omega (\mcc{L}/\omega)} 
 = - \frac{\mcc{L}_k}{\mcc{L}_\omega - \mcc{L}/\omega}
 = v_\varepsilon. 
\end{gather}
However, this generalization is arbitrary, and other definitions, such as $v_g = v_\rho$ or $v_g = v_{\mc{I}}$, would be equally justified. [In fact, the latter would be more fundamental, because the ACT holds also in nonstationary medium, unlike the energy conservation law.] More consistently, $v_g$ is defined as the velocity of \textit{information}, \ie the velocity on characteristics \cite{ref:whitham65, book:whitham}. Below, we restate how it is calculated in the general nonlinear problem. 

To find characteristics of \Eqs{eq:la}-\eq{eq:actwith}, let us consider traveling-wave solutions, with the propagation velocity $v_g$ yet to be found. In other words, let us search for solutions in the form where all the wave variables are expressed through a single variable $\zeta(x,t) \equiv x - X(t)$, such that $d_t X = v_g$; then, $\pd_x = d_\zeta$ and $\pd_t = -v_g\,d_\zeta$. In particular, one thereby gets from \Eq{eq:cons} that
\begin{gather}\label{eq:nlinvgchar}
v_g = \omega'/k' \equiv d_k \omega,
\end{gather}
where primes denote $d_\zeta$, and $d_k$ is taken in the sense that $\omega = \omega(k, a(k))$. [Notice that the latter local relation holds on a characteristic only, and generally there are two branches of $\omega(k, a(k))$ corresponding to the two different types of characteristics that we will find.] Hence, understanding the nonlinear group velocity as the characteristic velocity represents a natural generalization of~$v_{g0}$.

To actually find $v_g$, we proceed as follows. Using that
\begin{gather}
\pd_t \mcc{L}_\omega = \mcc{L}_{\omega a}\, \pd_t a + \mcc{L}_{\omega\omega}\, \pd_t \omega + \mcc{L}_{\omega k}\,\pd_t k, \\
\pd_x \mcc{L}_k = \mcc{L}_{k a}\, \pd_x a + \mcc{L}_{kk}\, \pd_x k + \mcc{L}_{k \omega}\,\pd_x \omega, 
\end{gather}
one can rewrite the ACT as
\begin{multline}\label{eq:c1}
v_g (\mcc{L}_{\omega a} a' + \mcc{L}_{\omega\omega} \omega' + \mcc{L}_{\omega k} k') + \\
\mcc{L}_{k a} a' + \mcc{L}_{kk} k' + \mcc{L}_{k \omega} \omega' = 0.
\end{multline}
Here $a'$ can be derived from \Eq{eq:la}, after differentiating the latter with respect to $\zeta$:
\begin{gather}
0 = d_\zeta \mcc{L}_a = \mcc{L}_{a a}\, a' + \mcc{L}_{a\omega} \omega' + \mcc{L}_{a k} k'.
\end{gather}
Specifically, one gets
\begin{gather}
\frac{da}{dk} = - \frac{\mcc{L}_{a\omega} v_g + \mcc{L}_{a k}}{\mcc{L}_{a a}},
\end{gather}
so \Eq{eq:c1} rewrites as follows
\begin{gather}\label{eq:quadr}
p v_g^2 + 2r v_g + q = 0,
\end{gather}
where we introduced
\begin{gather}
p = \mcc{L}_{a a}\mcc{L}_{\omega\omega} - \mcc{L}_{\omega a}^2, \\
r = \mcc{L}_{a a}\mcc{L}_{\omega k} - \mcc{L}_{\omega a}\mcc{L}_{ka}, \\
q = \mcc{L}_{kk}\mcc{L}_{a a} - \mcc{L}_{k a}^2.\label{eq:q}
\end{gather}

Since \Eq{eq:quadr} is a quadratic equation for $v_g$, there are generally \textit{two} group velocities different from each other (regardless of the type of nonlinearity), causing signal splitting. The exception is the linear regime. In that case, it is convenient to use $A$ instead of $a$ (\Sec{app:sflow}); then the same equations hold, if $\pd_a$ is replaced with $\pd_A$. On the other hand, $\mcc{L}_{AA} = 0$, so one obtains
\begin{gather}
v_g^2 \mcc{L}_{\omega A}^2 + 2v_g \mcc{L}_{\omega A}\mcc{L}_{kA} + \mcc{L}_{k A}^2 = 0
\end{gather}
(cf. also \Ref{book:landau6}), yielding that the two roots coincide:
\begin{gather}
v_g = - \mcc{L}_{kA}/\mcc{L}_{\omega A} = - \mcc{L}_{k}/\mcc{L}_{\omega} = v_{g0}. 
\end{gather}
Thus, we again see that the nonlinear $v_g$, defined as the characteristic velocity, equals $v_{g0}$ in the linear limit.

%-----------------------------------------------------------
\subsection{Stability criterion}
\label{sec:light}

If $r^2 < pq$, there are no real solutions for $v_g$, so no stable envelope is possible in this regime. This means that amplitude modulations will grow with time, \ie the wave is modulationally unstable, unless $r^2 \ge pq$. Below, we will put this criterion in a yet different form \cite{book:whitham}, which we will need in the main text. 

Let us choose the action density $\mc{I} \equiv \mcc{L}_\omega$ to serve as an independent variable instead of $a$. Also, using the NDR, exclude $\omega$ from the list of independent variables; hence,
\begin{gather}
\omega = \omega(k ,\mc{I}), \quad
a = a(k, \mc{I}), \quad
\mc{J} = \mc{J}(k, \mc{I}),
\end{gather}
where $\mc{J} \equiv - \mcc{L}_k$ is the action flux density. In particular, notice that $\mcc{L}$ takes the following form:
\begin{gather}
\mcc{L}(a(k,\mc{I}), \omega(k,\mc{I}), k) \equiv \Lambda(k, \mc{I}),
\end{gather}
so $\Lambda_k = \mc{I}\omega_k - \mc{J}$ and $\Lambda_\mc{I} = \mc{I}\omega_\mc{I}$, due to \Eq{eq:la}. Now \Eqs{eq:cons} and \eq{eq:actwith} on characteristics read~as
\begin{gather}
-v_g k' + \omega_k k' + \omega_{\mc{I}} \mc{I}' = 0, \quad
-v_g \mc{I}' + \mc{J}_k k' + \mc{J}_{\mc{I}} \mc{I}' = 0. \notag
\end{gather}
From $\Lambda_{k\mc{I}} = \Lambda_{\mc{I}k}$, it follows that $\omega_k = \mc{J}_{\mc{I}}$, so one~gets
\begin{gather}
v_g = \omega_k \pm \sqrt{\omega_{\mc{I}}\mc{J}_k}.
\end{gather}
(In a linear wave, $\omega_{\mc{I}} = 0$, so there is only one group velocity, $v_g = v_{g0}$.) Hence, the wave is stable if 
\begin{gather}\label{eq:lcrit}
\omega_{\mc{I}}\mc{J}_k > 0.
\end{gather}

In particular, when $\omega_{\mc{I}}$ is small, one can substitute in \Eq{eq:lcrit} the lowest-order approximation for $\mc{J}$, yielding $\omega_{\mc{I}}\mc{J}_k = \omega_{\mc{I}}\mc{I}\,\pd_k v_{g0}$. Suppose that, approximately, $\mc{I} \propto a^2$ (corresponding to the most common choice of $a$), so $\omega_{\mc{I}}\mc{I} = \omega_a a/2$. Then, required for stability is the condition $\omega_a\,\pd_k v_{g0} > 0$. The latter is equivalent to that flowing from, \eg Eq.~(4) in \Ref{ref:dewar72d} or Eq.~(7) in \Ref{ref:zakharov09} and also agrees with the Lighthill's original criterion for weakly nonlinear waves~\cite{ref:lighthill67}. Hence, in the main text, \Eq{eq:lcrit} is called the generalized Lighthill's criterion.

\gap
%%%%%%%%%%%%%%%%%%%%%%%%%%%%%%%%%%%%%%%%%%%%%%%%%%%%%%%%%%%%
\section{TPMI as the adiabatic limit of SI}
\label{app:kras}

Let us compare the TPMI rate that we found in \Sec{sec:pulse} with that flowing from \Ref{ref:krasovsky94} for the more general SI \cite{ref:kruer69, ref:goldman70, ref:goldman71, ref:brunner04}. Specifically, we will consider the limit $S \gg 1$, where Eq.~(23) of \Ref{ref:krasovsky94} applies, reading~as
\begin{gather}\label{eq:kr}
1 - \frac{\mc{A}}{\Delta \bar{\omega}^2} - \frac{(S \mc{A} \mc{T})^2}{[\Delta \bar{\omega} + \Delta \bar{k}(1 - \mc{T})]^2} = 0.
\end{gather}
Here $\mc{A} = \Omega_E^2$, $\mc{T} = 3\kappa^2$, $\Delta \bar{\omega} = (\Delta\omega - \Delta k\,u_0)/\omega_0$, $u_0 = \omega_0/k_0$, $\Delta \bar{k} = \Delta k/k_0$, and $S$ is defined as in our paper. 

Equation \eq{eq:kr} describes four eigenmodes, two of which correspond to oscillations at $\Delta \bar{\omega} \approx \pm \mc{A}^{1/2}$ and two others have yet lower frequencies. We will assume that the former are of zero amplitudes (remember that our theory applies only at time scales large compared to $\omega_E^{-1}$; see Paper~I), so those are the lower-frequency modes that we will consider. Hence we take $\Delta \bar{\omega}^2 \ll \mc{A}$, in which case the first term in \Eq{eq:kr} can be neglected, yielding
\begin{gather}
[\Delta \bar{\omega} + \Delta \bar{k}(1 - \mc{T})]^2 = - \frac{\Delta \bar{\omega}^2}{\mc{A}}\,(S \mc{A} \mc{T})^2.
\end{gather}
Treating the right-hand side as a perturbation, one gets
\begin{gather}
\Delta \bar{\omega} \approx \Delta \bar{k}(\mc{T} - 1) \pm i \Delta \bar{k} S \mc{T}\sqrt{\mc{A}}
\end{gather}
(here we also used that $\mc{T} \ll 1$), which is equivalent to
\begin{gather}
\Delta \omega \approx \Delta k\, v_{g0} \pm i (\Delta k/k)\, \omega_p S \mc{T}\sqrt{\mc{A}}.
\end{gather}
Since $\mc{T}\mc{A}^{1/2} = 3\Omega_E \kappa^2$, the instability rate $\gamma$ then equals
\begin{gather}
\gamma \approx \Delta k\,v_{g0}\Omega_E S,
\end{gather}
which precisely matches our \Eq{eq:gamma1} taken at $S \gg 1$.

%\bibliography{main,foot}

\begin{thebibliography}{10}
\bibitem{foot:xstar}
From $S \sim 1$, one gets another estimate, $\phi \sim (\nu/\kappa^3)^{1/2}$.
  Thus, $\kappa_* \sim \phi_0^{-1}(\nu/\kappa_0)^{1/2}$, and $\phi_* \sim
  \phi_0^{3/2}\kappa_0^{3/4}\nu^{-1/4}$.
\bibitem{foot:know}
To our knowledge, the results of \Ref{ref:krasovsky94} have never been
  associated with the TPMI before.
\bibitem{foot:nlse}
Unlike the linear Schr\"odinger equation (LSE), the NLSE cannot be obtained via
  inverse Fourier transform of the dispersion relation, because the NDR does
  not describe individual harmonics of the wave but rather the local $(a,
  \omega, k)$. Hence, an additional assumption is needed to add a nonlinear
  term to LSE. Usually postulated is that the wave energy flux is approximately
  quadratic in the amplitude \cite{book:karpman} (\ie that
  $\mcc{L} \propto a^2$ to the lowest order in $a$) or that the nonlinear term
  is a local function of $a$ \cite{ref:dewar72d} (rather than of $\omega$ or
  $k$). Clearly, these assumptions do not hold for waves with trapped
  particles.
\end{thebibliography}

\end{document}